\documentclass[iop, twocolumn]{aastex6}
\usepackage{amsmath}     
\usepackage{wasysym}           
\usepackage{graphicx}
\usepackage{amssymb}
\usepackage{epstopdf}
\usepackage{mathrsfs}
\usepackage{natbib}
\usepackage{color}
\usepackage{lipsum}

\begin{document}
\title{The Radiation Hydrodynamics of Relativistic Shear Flows}
\shorttitle{Relativistic shear flows}
\author{Eric R. Coughlin\altaffilmark{1}, Mitchell C. Begelman\altaffilmark{1}}
\shortauthors{\sc{Coughlin \& Begelman}} 
\email{eric.coughlin@colorado.edu, mitch@jila.colorado.edu}
\affiliation{JILA, University of Colorado and NIST, UCB 440, Boulder, CO 80309}
\altaffiltext{1}{Department of Astrophysical and Planetary Sciences, University of Colorado, UCB 391, Boulder, CO 80309}

\begin{abstract}
We present a method for analyzing the interaction between radiation and matter in regions of intense, relativistic shear that can arise in many astrophysical situations. We show that there is a simple velocity profile that should be manifested in regions of large shear that have ``lost memory'' of their boundary conditions, and we use this self-similar velocity profile to construct the surface of last scattering, or $\tau \simeq 1$ surface, as viewed from any comoving point within the flow. We demonstrate that a simple treatment of scattering from this $\tau \simeq 1$ surface exactly conserves photon number, and derive the rate at which the radiation field is heated due to the shear present in the flow. The components of the comoving radiation energy-momentum tensor are calculated, and we show that they have relatively simple, approximate forms that interpolate between the viscous (small shear) and streaming (large shear) limits. We put our expression for the energy-momentum tensor in a covariant form that does not depend on the explicit velocity profile within the fluid and, therefore, represents a natural means for analyzing general, radiation-dominated, relativistic shear flows.
\end{abstract}

\section{Introduction}
Radiation interacts dynamically with matter in many astrophysical systems. Often, the density of the plasma in these systems is high enough that the photon mean free path is very small compared to scales characterizing the fluid. In these optically thick situations, the radiation field is very nearly isotropic in the rest frame of the fluid, and the dynamical coupling is determined by local quantities. On the other hand, there are situations in which the mean free path of a photon greatly exceeds the physical length scale of the gas that scatters those photons. For these optically thin scenarios, the properties of the radiation field, and its dynamical effects, are largely determined nonlocally.

In these optically thick and thin limits, the manner in which radiation influences the gas is well-understood. For the former, the radiation field generates an energy density and pressure (in addition to the gas pressure) that are characterized by isotropy in the local comoving frame of the matter and an equation of state that reflects the relativistic nature of a photon gas (so that the pressure is one third the energy density; \citealt{mih84}). For the latter, the radiation field does not respond locally to the properties of the gas, but can be considered to be imposed externally. The angular distribution of radiation coupled with the state of motion of the gas then generates a ``radiation drag'' that can cause particles in the Solar System to gradually spiral into the Sun (the Poynting-Robertson effect; \citealt{rob37}) or an acceleration that can propel material away from active galactic nuclei (the Compton rocket effect; \citealt{ode81, phi82}). 

An interesting question then arises when one considers how radiation couples to a gas that is between the optically thick and thin limits. Some progress has been made on this question from the optically thick standpoint by assuming that the radiation field is \emph{approximately} isotropic in the instantaneous rest frame of the fluid. The anisotropic contribution can then be determined from the Boltzmann equation, yielding the general relativistic equations of radiation hydrodynamics in the viscous limit \citep{cou14b}. These equations are ``viscous'' in the sense that the shear of the fluid -- the change in the fluid velocity over the mean free path of a photon -- serves to transfer energy and momentum between the radiation and the scatterers. However, it is not at all apparent how this viscous coupling in the optically thick limit transitions to the optically thin counterpart of radiation drag, where the relevant length scale changes from the (assumed-small) mean free path of the photon to the size of the entire fluid (or longer).

This question is also of practical, not just theoretical, importance, as some astrophysical systems are characterized by  regions of only marginal optical depth. Such a region naturally arises when an optically thin jet or wind propagates alongside an optically thick disk or envelope, which occurs in ultraluminous X-ray sources (e.g., \citealt{ara92, beg06}), the collapsar model of long gamma-ray bursts (\citealt{woo93, mac99}), and super-Eddington tidal disruption events \citep{cou14a}, to name a few. The transition between the jet and the envelope is then of marginal optical depth, and one must self-consistently account for both the evolution of the fluid and the radiation field within this transition. 

Numerically, most authors have resorted to some form of ``closure'' to capture the important physical effects that take place in these regions of marginal optical depth (but see \citealt{jia14} and \citealt{rya15}), popular choices being flux-limited diffusion (FLD; \citealt{lev81, ohs09}) and M1 \citep{lev84, mck14}. However, FLD does not capture any relativistic or velocity-dependent effects, which can be quite large in a region of high velocity and intense shear such as the transition between a fast-moving outflow and a hydrostatic envelope. An advantage of M1 is that it is easily extended to incorporate all of the effects of general and special relativity \citep{sad13}. However, M1 breaks down when the radiation field is inherently very anisotropic (for example, along the axis of the jet as it propagates alongside the envelope) and it does not reduce correctly to the viscous limit (one must apply an additional, viscous term to the stress tensor in the radiation-rest frame: \citealt{cou14b, sad14}). 

In this paper we pursue a simple, but physically-motivated approach to analyzing the manner by which scatterers in a fast-moving fluid interact with radiation when the medium is optically thick, thin, or in-between, with the specific application of the intense shear layers generated between relativistic jets and their surroundings in mind. In particular, we assume that when such a shear layer develops, the fluid rapidly assumes a self-similar form so that the fluid looks identical at every comoving point. In Section 2 we show that this assumption of self-similarity yields a very specific, one-parameter family of velocity profiles that is described solely by the amount of shear present in the flow. 

We further assume that the photons interacting with a given fluid element were all scattered on a prescribed surface located at an optical depth $\tau \simeq 1$ relative to that fluid element, and that the scattering is isotropic in the rest frame of the scatterer and elastic (i.e., the scattering takes place in the Thomson limit). We calculate the shape of the $\tau = 1$ surface in Section 3, and we show how it depends on the amount of shear within the flow.

Section 4 combines the assumption of the self-similarity of the flow and the $\tau = 1$ scattering specification, demonstrating that the number density of photons is manifestly conserved, both in time and space, throughout the shear layer (as must be true, since we are only considering photon scattering). We also show that, in order for the radiation energy density to be simultaneously conserved in a scattering event and uniform across the shear layer, both necessary for the consistency of this model, the photon energy density must be an increasing function of time and thereby heating up in response to the shear present within the flow. In section 5 we solve for this heating rate in the optically thin and thick limits, and we provide an approximate, interpolated solution for the heating rate when the optical depth is marginal. Section 6 presents the radiation energy-momentum tensor, and we derive the functional forms of the shear stress and the pressures of the radiation field, showing that they have relatively simple, approximate, analytic forms that match the viscous and streaming limits. We summarize and discuss the implications of our findings in Section 7.

\section{Velocity field}
Consider a two-dimensional, planar fluid where the fluid motion is purely along $z$ and the variation in that motion is purely along $y$. The three-velocity of this fluid is characterized by $\mathbf{v} = v(y)\hat{z}$, and the four-velocity is, correspondingly,

\begin{equation}
U^{\alpha} = \left(
\begin{array}{c}
\Gamma \\
0 \\
0 \\
\Gamma{v}
\end{array} \right),
\end{equation}
where $\Gamma = (1-v^2)^{-1/2}$ is the Lorentz factor. 

The fundamental assumption we will make here is that the fluid appears identical to every comoving observer. This is essentially a statement of the self-similarity of the flow, and we suggest that it may apply in regions of shear that have ``lost memory'' of their boundary or initial conditions. This assumption then implies that the comoving density of matter satisfies $\rho'(y) = \rho'$, with $\rho'$ a constant, and that the velocity field varies as

\begin{equation}
\Gamma = \cosh(\mu\tau'_y),\quad \Gamma{v} = \sinh(\mu\tau'_y), \label{gammacosh}
\end{equation}
where

\begin{equation}
\tau'_{y} = \int_0^{y}\rho'\,\kappa\,d\tilde{y} = \rho'\kappa\,{y} \label{tauy}
\end{equation}
is the comoving optical depth along the $y$ direction ($\kappa$ is the opacity) and $\mu$ is a constant that describes the amount of shear present in the flow. We can see that this expression for $v$ has the required properties by considering a fluid element within the flow that moves with four-velocity $U^{\alpha}$ with respect to some frame $\alpha$; relative to another fluid parcel within the shear layer, the frame of which we will denote by $\beta$, that same fluid element is observed to be moving at a velocity $U^{\beta}$ that is related to $U^{\alpha}$ by

\begin{equation}
U^{\beta} = \Lambda^{\beta}_{\,\,\alpha}U^{\alpha},
\end{equation}
where $\Lambda^{\beta}_{\,\,\alpha}$ is the local Lorentz transformation between the $\alpha$ and $\beta$ frames. We thus have

\begin{equation}
\Lambda^{\beta}_{\,\,\alpha} = \left(
\begin{array}{cccc}
\Gamma_1 & 0 & 0 & -\Gamma_1v_1 \\
0 & 1 & 0 & 0\\
0 & 0 & 1 & 0\\
-\Gamma_1v_1 & 0 & 0 & \Gamma_1
\end{array} \right),
\end{equation}
where $v_1$ is the three-velocity of the $\beta$ frame with respect to the $\alpha$ frame, and 

\begin{equation}
U^{\alpha} = \left(
\begin{array}{c}
\Gamma_2 \\
0 \\
0 \\
\Gamma_2v_2
\end{array} \right),
\end{equation}
where $v_2$ is the three-velocity of the fluid parcel as measured in the $\alpha$ frame. If we now use equation \eqref{gammacosh} for the velocities, then the velocity of the fluid parcel as measured in the $\beta$ frame is

\begin{equation}
U^{\beta} = \left(
\begin{array}{c}
\cosh(\mu\tau'_1)\cosh(\mu\tau'_2)-\sinh(\mu\tau'_1)\sinh(\mu\tau'_2) \\
0 \\
0 \\
-\sinh(\mu\tau'_1)\cosh(\mu\tau'_2)+\cosh(\mu\tau'_1)\sinh(\mu\tau'_2)
\end{array} \right)
\end{equation}
\begin{equation*}
= \left(
\begin{array}{c}
\cosh[\mu(\tau'_2-\tau'_1)] \\
0 \\
0 \\
\sinh[\mu(\tau'_2-\tau'_1)]
\end{array}\right),
\end{equation*}
where $\tau'_1$ is the optical depth of the $\beta$ frame and $\tau'_2$ is the optical depth of the moving gas parcel, both with respect to the $\alpha$ frame. However, it is apparent that

\begin{equation}
\tau'_2 - \tau'_1 = \int_0^{y_2}\rho'\,\kappa\,d\tilde{y} - \int_0^{y_1}\rho'\,\kappa\,d\tilde{y}
\end{equation}
\begin{equation*}
= \int_{y_1}^{y_2}\rho'\,\kappa\,d\tilde{y},
\end{equation*}
where $y_1$ and $y_2$ are the positions of the $\beta$ frame and the gas parcel, respectively, as measured by the $\alpha$ frame. This expression is just the optical depth to the gas parcel as measured by the $\beta$ frame, and we thus see that the velocity field as measured in the $\beta$ frame is identical to the velocity field measured in the $\alpha$ frame -- precisely the attribute we require for the flow.

Interestingly, this form for the velocity field, equation \eqref{gammacosh}, also describes the ultrarelativistic limit of optically thick jet propagation mediated by radiation viscosity. We refer the reader to the Appendix for a demonstration of this result. 

\section{$\tau=1$ surface}
According to a given fluid element, which we will consider to be the origin, neighboring gas parcels are all observed to have Doppler-shifted volumes of

\begin{equation}
V_o = \frac{V'}{\Gamma(1+v\cos\theta)}. \label{Vo}
\end{equation}
In this equation, $V_o$ is the observed volume of the gas parcel, $V'$ is its comoving volume, $v$ is its velocity, and $\theta$ is the angle made between the $z$-axis of the origin and the velocity vector of the moving gas parcel and is, therefore, identical to the normal definition of $\theta$ in spherical-polar coordinates. 

The perceived optical depth measured to a distance $r$ from the origin is given by

\begin{equation}
\tau = \int_0^{r}\rho_o\,\kappa\,{d\tilde{r}}, \label{tau1}
\end{equation}
where $\tilde{r}$ is a dummy variable of integration and $\rho_o$ is the observed mass density within the fluid; note that this is different from the comoving optical depth, defined by equation \eqref{tauy}, as here we are taking into account light-travel time and Lorentz contraction effects (which are encapsulated in the Doppler factor). Since the observed density is related to the observed volume via $\rho_o \propto 1/V_{o}$, equation \eqref{tau1} becomes

\begin{equation}
\tau = \int_{0}^{r}\Gamma\left(1+v\cos\theta\right)\rho'\,\kappa\,d\tilde{r}. \label{taur}
\end{equation}

With equation \eqref{gammacosh} for the velocity, equation \eqref{taur} is

\begin{equation}
\tau = \frac{1}{\lambda'}\int_0^{r}\left[\cosh\left(\frac{\mu{y}}{\lambda'}\right)+\sinh\left(\frac{\mu{y}}{\lambda'}\right)\cos\theta\right]d\tilde{r},
\end{equation}
where $\lambda' = 1/(\rho'\kappa)$ is the comoving mean free path of the radiation. If we further note that our definition of $\theta$ is just that of spherical-polar coordinates, so we can write $y = r\sin\theta\sin\phi$, then this integral can be evaluated and inverted to yield $r$ in terms of $\tau$, $\mu$, and the polar angles. Doing so gives

\begin{equation}
r = \frac{\lambda'}{\mu\sin\theta\sin\phi}\ln\left(h\right), \label{rsurf}
\end{equation}
where

\begin{multline}
h = \left(1+\cos\theta\right)^{-1}\bigg{(}\cos\theta+\mu\tau\sin\theta\sin\phi \\ +\sqrt{1+2\mu\tau\sin\theta\cos\theta\sin\phi+\mu^2\tau^2\sin^2\theta\sin^2\phi}\bigg{)} \label{heq}
\end{multline}
is a function of $\mu$, $\tau$ and the polar angles. Equation \eqref{rsurf} gives the perceived distance to neighboring fluid elements in terms of their optical depth $\tau$, which is a two-dimensional surface in $\theta$ and $\phi$ for fixed $\tau$. 

We expect that the radiation field at the origin is mainly determined by photons scattered at $\tau \simeq 1$. Photons scattered at $\tau \gtrsim1$ will, on average, suffer additional scattering before reaching the origin, while relatively few photons will be scattered at $\tau \lesssim 1$. 

\begin{figure*}[htbp] 
   \centering
   \includegraphics[width=\textwidth]{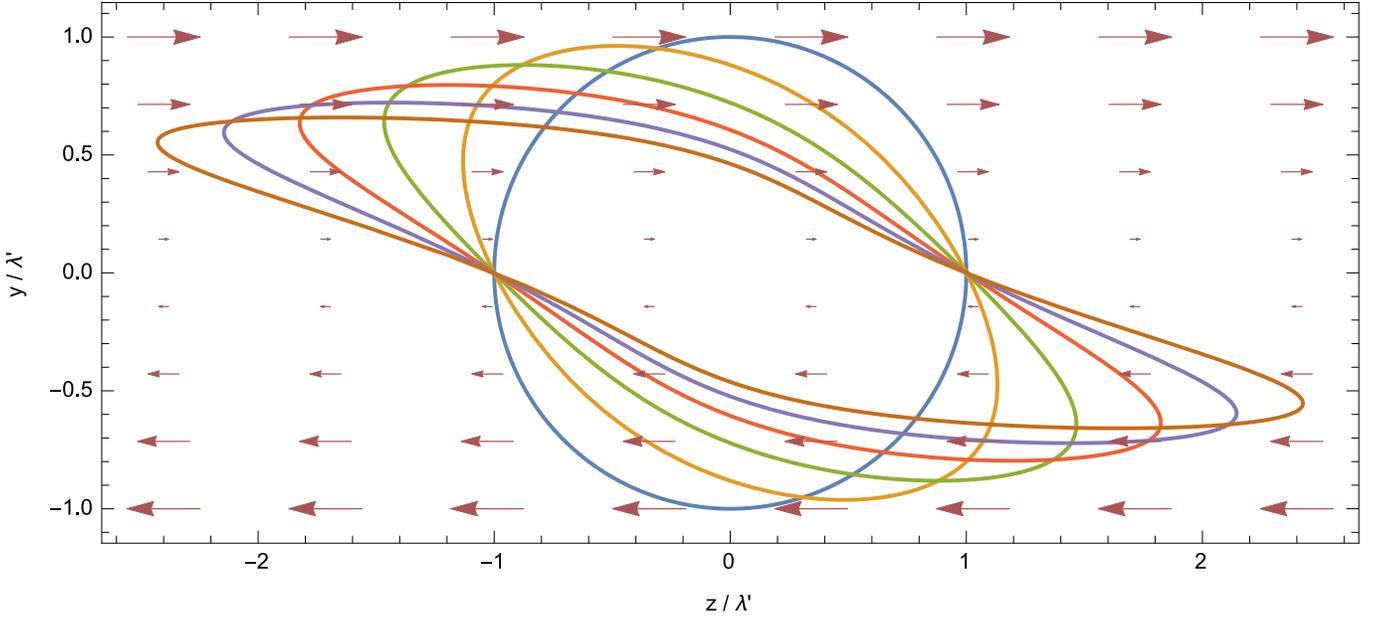} 
   \caption{The $\tau=1$ surface in the $y-z$ plane for $\mu =0$ -- 5 in increments of 1; the circular surface corresponds to $\mu = 0$, while the most elongated surface corresponds to $\mu = 5$. The arrows serve to indicate the direction of motion of the fluid, and the length of the arrow gives an indication of its magnitude.}
   \label{fig:tausurfs}
\end{figure*}

Figure \ref{fig:tausurfs} shows the $\tau = 1$ surface in the $y-z$ plane for a number of different $\mu$, which we recall parameterizes the change in the velocity in the $y$-direction relative to the optical depth, with larger $\mu$ corresponding to greater shear (see Equation \ref{gammacosh}). The blue circle is the solution for $\mu = 0$, the most elongated surface has $\mu = 5$, and each curve in between differs from the previous by 1. The arrows in the figure show the direction of the velocity, with the length of the arrow scaling directly with the magnitude of the velocity. 

This figure demonstrates how the optical depth of the fluid responds to changes in the velocity field: when $\mu = 0$, the fluid is stationary everywhere, and correspondingly the $\tau = 1$ surface is just a circle where $r = \lambda'$. As the shear starts to increase, the fluid elements that are directly along $y$ at $z = 0$ are Lorentz contracted because their motion is perpendicular to the line of sight, which increases the density of those fluid elements and brings the $\tau=1$ surface closer to the origin. As we look along $y$ at non-zero locations on the $z$-axis, the Doppler shift competes with the Lorentz contraction to give a more complicated surface. In particular, for $y < 0$ and $z < 0$, the fluid is moving away from the origin, and the Doppler shift works with the Lorentz contraction to give an overall smaller fluid volume, increasing the density and moving the $\tau = 1$ surface closer to the origin. On the other hand, for $y > 0$ and $z < 0$, the fluid is moving towards the origin, and the Doppler shift serves to lengthen the fluid element, decreasing the optical depth and extending the $\tau=1$ surface to greater distances from the origin. This behavior is inverted when we consider positions within the fluid characterized by $z > 0$. 

\begin{figure}[htbp]
   \centering
   \includegraphics[width=0.5\textwidth]{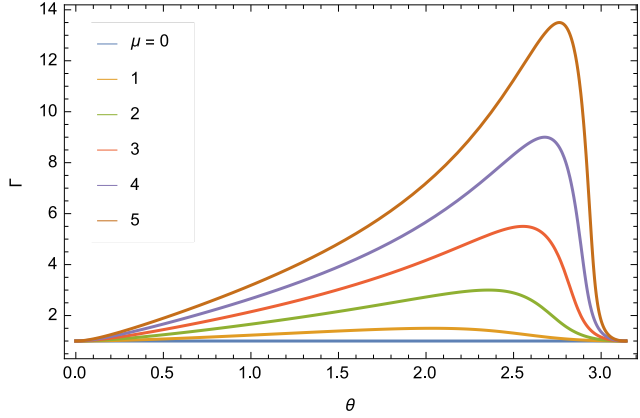} 
   \caption{The Lorentz factor as a function of $\theta$ on the $\tau = 1$ surface for the same values of $\mu$ chosen in Figure \ref{fig:tausurfs}. Here we chose $\phi = \pi/2$, for which the maximum in the Lorentz factor is achieved near $\theta \simeq \pi-1/\mu$. }
   \label{fig:gammaofmu}
\end{figure}

Using equations \eqref{gammacosh} and \eqref{rsurf}, we find that the fluid four-velocity varies along the $\tau = 1$ surface as

\begin{equation}
\Gamma = \frac{1}{2}\left(h+\frac{1}{h}\right), \label{gammaofh}
\end{equation}
\begin{equation}
\Gamma{v} = \frac{1}{2}\left(h-\frac{1}{h}\right). \label{gammavofh}
\end{equation}
where $h$ depends on $\theta$ and $\phi$ via equation \eqref{heq}. Figure \ref{fig:gammaofmu} shows the Lorentz factor -- equation \eqref{gammaofh} -- along the $\tau=1$ surface with $\phi = \pi/2$ for the same set of $\mu$ used in Figure \ref{fig:tausurfs}. This figure shows that the Lorentz factor, and correspondingly the velocity, is maximized approximately at the location where the $\tau=1$ surface has receded to the largest distance from the origin, which is consistent with the fact that the Doppler shift is modifying the observed mean free path of the radiation. 
We will use these expressions in the following section.

\section{Conserved quantities}
\subsection{Photon number}
Because we are only considering scattering processes, the \emph{number} of photons must be conserved and, to preserve the similarity of the flow, must also be the same at every point within the fluid. On the $\tau\simeq1$ surface, we have

\begin{equation}
n' = \int{f\cdot({k'})^2dk'\,d\Omega'}, \label{np0}
\end{equation}
where $f$ is the photon distribution function (units of number per momentum cubed per length cubed; see, e.g., \citealt{mih84}), $k'$ is the magnitude of the photon three-momentum, and $\Omega'$ is the comoving solid angle. A prime on a quantity denotes that it is measured in the comoving frame of the $\tau\simeq1$ surface. We will adopt the assumption that photons are scattered isotropically in the rest frame of the fluid, and so the distribution function should be independent of angle. It is thus tempting to write $f = f(k')$; however, this assertion is problematic as the radiation field should be heating up if there is a non-zero shear ($\mu \neq 0$), and we therefore expect the photon spectrum to be an increasing function of time. We will therefore adopt the expression

\begin{equation}
f = g(t)\,q\left(\frac{k'}{j(t)}\right), \label{dist1}
\end{equation}
where $g$ and $j$ are functions of time and $q$ is an unspecified function. Inserting this expression into equation \eqref{np0}, we find

\begin{equation}
n' = 4\pi{}g(t)j(t)^3\int{}q(x)x^2dx, \label{np1}
\end{equation}
where we have changed variables to $x \equiv k'/j(t)$. If we now enforce the fact that the photon number density be time-independent (scattering does not alter the number of photons), then we immediately see that

\begin{equation}
g(t) = j(t)^{-3}.
\end{equation}

The number density of photons at the origin is

\begin{equation}
n = \int{}f_o\,k^2dk\,d\Omega, \label{n0}
\end{equation}
where $f_o$ is the distribution function at the origin. Since the distribution function is a general relativistic invariant \citep{deb09a}, we have that the distribution function observed at the origin is the same as that on the scattering surface:

\begin{equation}
f_o = j(t_r)^{-3}q\left(\frac{k'}{j(t_r)}\right), \label{fo}
\end{equation}
where $t_r = t-r(\theta,\phi)$ is the \emph{retarded time}; $r$ is the distance to the $\tau = 1$ surface that is a function of angle, specifically given by equation \eqref{rsurf}. Inserting this expression into equation \eqref{n0} gives

\begin{equation}
n = \int{}q\left(\frac{k'}{j(t_r)}\right)\frac{k^2}{j(t_r)}d\left(\frac{k}{j(t_r)}\right)d\Omega. \label{n1}
\end{equation}
Because the photon four-momentum transforms as a vector, we have\footnote{The plus sign in this equation comes from the fact that the angle of the photon emitted by the $\tau \simeq 1$ surface has a direction that is offset from the angular location of the $\tau \simeq 1$ surface itself; thus, if a photon is coming from the $\{-z,y\}$ direction, its momentum is in the $\{z,-y\}$ direction, etc.}

\begin{equation}
k' = \Gamma{k}\left(1+v\cos\theta\right), \label{kpk}
\end{equation}
and so equation \eqref{n1} becomes

\begin{equation}
n = \frac{n'}{4\pi}\int\mathscr{D}^3d\Omega,
\end{equation}
where

\begin{equation}
\mathscr{D} = \frac{1}{\Gamma\left(1+v\cos\theta\right)}
\end{equation}
is the relativistic Doppler factor -- the same one that appears in equation \eqref{Vo}. By using equations \eqref{heq}, \eqref{gammaofh}, and \eqref{gammavofh}, we can show that the Doppler factor is given by

\begin{equation}
\mathscr{D} = \frac{1}{\sqrt{1+2\mu\tau\sin\theta\cos\theta\sin\phi+\mu^2\tau^2\sin^2\theta\sin^2\phi}}, \label{Doppler}
\end{equation}
and so we have

\begin{equation}
n = \frac{n'}{4\pi}\int\frac{\sin\theta\,{d\theta}\,d\phi}{\left(1+2\mu\tau\sin\theta\cos\theta\sin\phi+\mu^2\tau^2\sin^2\theta\sin^2\phi\right)^{3/2}}.
\end{equation}
Now, to preserve the self-similarity of the flow, the origin can equally as well be considered to be located on the $\tau=1$ surface appropriate to some other region of the fluid, and hence the number of emitted photons must be equal to the number of observed photons. We therefore require that $n = n'$, and the consistency of this approach then demands that

\begin{equation}
\int\frac{\sin\theta\,{d\theta}\,d\phi}{\left(1+2\mu\tau\sin\theta\cos\theta\sin\phi+\mu^2\tau^2\sin^2\theta\sin^2\phi\right)^{3/2}} = 4\pi.
\end{equation}
Remarkably, we can show that the above condition is satisfied for \emph{any product $\mu\tau$}, and it is therefore an identity. This model, in which the velocity varies according to equation \eqref{gammacosh}, thus exactly conserves particle number (as a function of time) and yields precisely the same particle number at every comoving point within the flow.

\subsection{Photon energy}
The energy density of the radiation scattered by a given location on the $\tau=1$ surface can be written as

\begin{equation}
e' = \int{}f\cdot(k')^3dk'd\Omega' = 4\pi{}j(t)\int{}q(x)x^3dx, \label{ep0}
\end{equation}
where here we have simply made the same transformation that turned equation \eqref{np0} into \eqref{np1}. As was done for the number density of photons, we can also construct the energy density observed at the origin, and thereby scattered by the $\tau=1$ surface, by

\begin{equation}
e = \int{}f\,k^3dk\,d\Omega.
\end{equation}
By now using the same procedure that allowed us to arrive at equation \eqref{n0}, we can show that this expression becomes

\begin{equation}
e = \frac{e'}{4\pi}\int{}\frac{j(t_r)}{j(t)}\mathscr{D}^4d\Omega.
\end{equation}
Because this is a scattering process that is assumed to be elastic in the rest frame of the scatterer, we require $e = e'$. This requirement then imposes the restriction

\begin{equation}
\int\frac{j(t_r)}{j(t)}\mathscr{D}^4d\Omega = 4\pi.
\end{equation}
Since this equation must be true for all time, we see that $j$ must have the form of an exponential:

\begin{equation}
j = \exp\left(\frac{\nu{t}}{\lambda'\tau}\right), \label{jexp}
\end{equation}
where $\nu$ is the dimensionless heating rate. We see that, in order for our treatment to be consistent, the heating rate must uniquely yield

\begin{equation}
\int\mathscr{D}^4e^{-\nu{r}/(\lambda'\tau)}d\Omega = 4\pi.
\end{equation}
Using equation \eqref{Doppler} for $\mathscr{D}$ and \eqref{rsurf} for $r$, this relation becomes

\begin{equation}
I(\nu,\mu\tau) = 4\pi, \label{nueq1}
\end{equation}
where 

\begin{equation}
I \equiv \int\frac{h^{-\frac{\nu}{\mu\tau\sin\theta\sin\phi}}\sin\theta\,d\theta\,d\phi}{\left(1+2\mu\tau\sin\theta\cos\theta\sin\phi+\mu^2\tau^2\sin^2\theta\sin^2\phi\right)^2} \label{Ieq}
\end{equation}
and $h$ is given by equation \eqref{heq}. 

\section{Heating rate}
From equation \eqref{ep0}, the energy density of the photon field evolves as $e' \sim \exp(\nu{t}/(\lambda'\tau))$, and $\nu$ therefore represents the rate at which the shear present in the flow provides energy to the radiation field. This heating rate is not arbitrary, however, as equation \eqref{nueq1} must be an identity for all $\mu\tau$ to preserve the self-similarity of the flow. Ideally we would like to solve equation \eqref{nueq1} for $\nu(\mu,\tau)$. However, the complexity of the integral $I$ makes this a formidable task, and we must resort to approximate solutions.

\subsection{Viscous limit}
When the product $\mu\tau$ is small, we can Taylor expand equation \eqref{nueq1} in powers of $\mu\tau$ about zero. Doing so to third order and evaluating the integrals, we can show that the heating rate must be

\begin{equation}
\nu = \frac{2}{15}\mu^2\tau^2+\mathcal{O}(\mu^4\tau^4) \label{nuvis}
\end{equation}
to satisfy equation \eqref{nueq1} identically (i.e., for all $\mu\tau$). This expression shows that the rate at which the radiation field heats viscously is proportional to the shear squared to lowest order, which is reasonable from a physical standpoint: if there is no shear ($\mu = 0$), we do not expect the radiation field to heat up. Likewise, the heating rate should not depend on the sign of $\mu$, and so we would predict that the lowest-order correction to the heating rate is proportional to $\mu^2$ (and all odd powers of $\mu$ should drop out of the expression).

\subsection{Streaming limit}
When the change in the Lorentz factor across the photon mean free path is much greater than one, so $\mu\tau \gg 1$, simply Taylor expanding $\nu$ about $\mu\tau=0$ is no longer a valid approach. In this case we must find an alternative method of approximating the integral $I$. 

To this end, note that when $\mu\tau \gg 1$, the denominator in equation \eqref{Ieq} is very large unless $\mu\tau\sin\theta\sin\phi \simeq 1$. Thus, it must be the case that $\mathscr{D}$ has a relative maximum near this location, and it is therefore this region of angular space that contributes predominantly to the integral $I$. By differentiating $\mathscr{D}$ with respect to $\theta$ and $\phi$, we find that the relative maxima occur at the points

\begin{equation}
\phi_m = \frac{\pi}{2},\frac{3\pi}{2},
\end{equation}
\begin{equation}
\cos\theta_m = \pm\sqrt{\frac{1}{2}\left(1+\frac{m}{\sqrt{4+m^2}}\right)} \simeq \pm\left(1-\frac{1}{2m^2}\right), \label{thm}
\end{equation}
where for ease of notation we have set $m \equiv \mu\tau$; in this expression the negative solution corresponds to $\phi = \pi/2$ and the positive solution to $\phi=3\pi/2$. We can show that the integrand is symmetric about $\phi \rightarrow \pi+\phi$ and $\theta \rightarrow \pi-\theta$, and for this reason we focus only on the maximum $\phi_m = 3\pi/2$. The last line in equation \eqref{thm} follows from the fact that we are interested in the large-$m$ limit of this solution, and this also shows that

\begin{equation}
\theta_m \simeq \frac{1}{m} + \mathcal{O}\left(\frac{1}{m^3}\right).
\end{equation}
With this value of $\theta_m$, we can approximate the function $\mathscr{D}$ by its second-order Taylor series about the point $\theta_m$. Calculating the second derivatives of $\mathscr{D}$, we can then show that this Taylor series is

\begin{equation}
\mathscr{D} \simeq m-\frac{m^5}{2}\left(\theta-\frac{1}{m}\right)^2-\frac{m}{2}\left(\phi-\frac{3\pi}{2}\right)^2.
\end{equation}
This expression shows that the angular width in the $\theta$ direction subtended by $\mathscr{D}$ about its maximum value is

\begin{equation}
\Delta{\theta}_m \simeq \frac{1}{m^2},
\end{equation}
while $\Delta\phi_m$ -- the angular width in the $\phi$ direction -- is of order unity. From this expression, it then follows that the angular integral of $\mathscr{D}$ is approximately

\begin{equation}
\int{}\mathscr{D}d\theta{d\phi} \simeq m\Delta\theta_m\Delta\phi_m \simeq \frac{1}{m}.
\end{equation}

We would like to follow a similar procedure for approximating the integral in equation \eqref{nueq1}, but the integrand $\mathscr{I} \equiv \mathscr{D}^4\exp{[-\nu{r}/(\lambda\tau)]}\sin\theta$ is clearly much more complicated than just $\mathscr{D}$. The maximum value of $\mathscr{I}$ will therefore not coincide exactly with the point $\theta_m = 1/m$. However, it is possible to show that the function

\begin{equation}
f \equiv \mathscr{D}^4\sin\theta
\end{equation}
has its maximum value at the point 

\begin{equation}
\theta_f \simeq 1/m+\mathcal{O}(1/m^3),
\end{equation}
while we find numerically that the function

\begin{equation}
g \equiv e^{-\frac{\nu{r}}{\lambda\tau}}
\end{equation}
has a relative minimum near the point $\theta \simeq 1/m$. It is also straightforward to show that the extrema of both of these functions occur at $\phi = \pi/2, 3\pi/2$. 

Since $\mathscr{I}$ is the product of the functions $f$ and $g$, the former possessing a relative maximum at the point $\theta_m \simeq 1/m$, the latter a relative minimum at that point, it is not immediately obvious that the relative extremum of $\mathscr{I}$ (which does occur at $\theta_m \simeq 1/m$) will be a maximum. However, we find that the absolute value of the second derivatives of $f$ at $\theta_m$ are much larger than the second derivatives of $g$. Therefore, the second derivatives of $\mathscr{I}$, which are just the sum of the second derivatives of $f$ and $g$, are dominated by the function $f$, meaning that the extremum of $I$ is indeed a maximum. We therefore have

\begin{equation}
\mathscr{I} \simeq \mathscr{I}_m+\frac{1}{2}\frac{\partial^2\mathscr{I}_m}{\partial\theta^2}\left(\theta-\frac{1}{m}\right)^2+\frac{1}{2}\frac{\partial^2\mathscr{I}_m}{\partial\phi^2}\left(\phi-\frac{3\pi}{2}\right)^2,
\end{equation}
where subscript $m$'s denote that we are evaluating the function at the maximum. We can show that the derivatives, to lowest order in $1/m$, are

\begin{equation}
\mathscr{I}_m \simeq m^{3-\nu},
\end{equation}
\begin{equation}
\frac{\partial^2\mathscr{I}_m}{\partial\theta^2} \simeq -m^{7-\nu},
\end{equation}
\begin{equation}
\frac{\partial^2\mathscr{I}_m}{\partial\phi^2}\simeq -m^{3-\nu},
\end{equation}
which demonstrates, as was true for $\mathscr{D}$ itself, that the width of the maximum of $\mathscr{I}$ is $\Delta\theta_m \simeq 1/m^2$ and $\Delta\phi_m \simeq 1$. It then follows that

\begin{equation}
I \simeq \int{\mathscr{I}d\theta{d\phi}} \simeq \mathscr{I}_m\Delta\theta_m\Delta\phi_m \simeq m^{1-\nu}.
\end{equation}

From equation \eqref{nueq1}, the integral $I$ must be equal to $4\pi$, so $\nu(m)$ must satisfy

\begin{equation}
m^{1-\nu} \simeq 4\pi.
\end{equation}
This shows that, in the large-$m$ limit, $\nu$ is given by

\begin{equation}
\nu = 1-\frac{C}{\ln(\mu\tau)},
\end{equation}
where $C$ is a numerical constant. Since we only took the leading-order (in $1/m$) expressions for $\mathscr{I}_m$ and its derivatives, the precise value of $C$ is not able to be directly computed here. This approach does show, however, that the heating rate \emph{must be equal to exactly one} in the limit that $\mu\tau \rightarrow \infty$, but the convergence is slow ($\propto 1/\ln(\mu\tau)$).

\subsection{Interpolated solution}
\begin{figure}[htbp]
   \centering
   \includegraphics[width=0.5\textwidth]{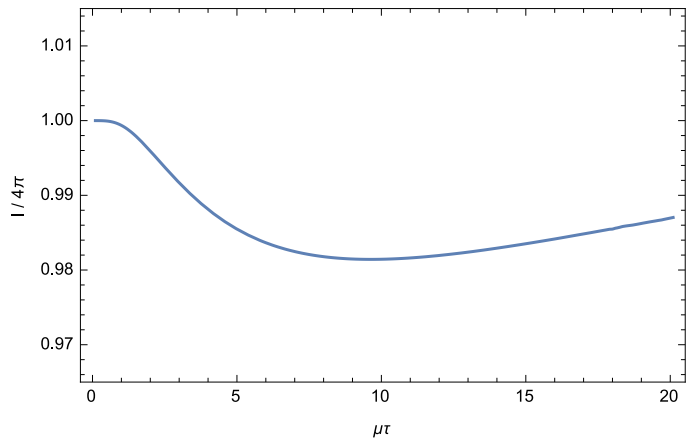} 
   \caption{The integral $I$, given by equation \eqref{Ieq}, normalized by $4\pi$ when the heating rate is given by equation \eqref{nueq2} with $C = 0.812$. This Figure demonstrates that this interpolated heating rate almost exactly solves the integral constraint \eqref{nueq1}.}
   \label{fig:Iofm}
\end{figure}

In the above two subsections we found the following asymptotic limits for the heating rate:

\begin{equation}
\nu = \begin{cases}
\frac{2}{15}\mu^2\tau^2\quad\quad\text{for }\mu\tau \ll 1 \\
1-\frac{C}{\ln(\mu\tau)}\,\,\,\text{        for   }\mu\tau \gg 1
\end{cases}
\end{equation}
To determine the $\mu$ dependence of $\nu$ in between these limits, we will write

\begin{equation}
\nu = 1-\frac{1}{1+\frac{1}{2C}\ln\left(\frac{P(\mu\tau)}{Q(\mu\tau)}\right)}. \label{nuapp}
\end{equation}
where $P$ and $Q$ are polynomials in the quantity $\mu\tau$. We can then determine these polynomials by requiring that equation \eqref{nuapp} reduce correctly to the viscous limit to a predetermined order. For example, if we only want to match the lowest-order viscous approximation to the heating rate, so $\nu = 2\mu^2\tau^2/15$, then we can show that 

\begin{equation}
P(\mu\tau) = 1+\frac{4C}{15}\mu^2\tau^2,
\end{equation}
and 

\begin{equation}
Q(\mu\tau) = 1.
\end{equation}
The expression for the heating rate that matches both the $\mu\tau \gg 1$ and $\mu\tau \ll 1$ limits is then

\begin{equation}
\nu = 1-\frac{1}{1+\frac{1}{2C}\ln\left(1+\frac{4C}{15}\mu^2\tau^2\right)}. \label{nueq2}
\end{equation}
To determine $C$ we have adopted a brute-force method of numerically integrating the left-hand side of equation \eqref{nueq1} with equation \eqref{nuapp} for $\nu$ for a number of different $C$. We find that the value of $C$ that solves equation \eqref{nueq1} when $\mu\tau \gg 1$ is

\begin{equation}
C \simeq 0.812.
\end{equation}
Figure \ref{fig:Iofm} shows the integral in equation \eqref{nueq1}, $I$, normalized by $4\pi$ when $\nu$ is given by equation \eqref{nueq2} and $C = 0.812$. It is apparent that this expression for the heating rate almost exactly satisfies the integral constraint \eqref{nueq1}, with the maximum deviation from unity being $0.98$ at $\mu\tau \simeq 10$. 

If we want to match higher-order viscous corrections, then it is apparent that $P$ and $Q$ will be of the general form

\begin{equation}
P(\mu\tau) = \sum_{n=0}^{j}p_n(\mu\tau)^{n},
\end{equation}
\begin{equation}
Q(\mu\tau) = \sum_{n=0}^{\ell}q_n(\mu\tau)^{n}.
\end{equation}
Because the heating rate should only depend on even powers of $\mu\tau$, it follows that the odd coefficients in these expansions are zero. Likewise, since the asymptotic limit should be $\sim 1-C/\ln(\mu\tau)$, we find $j = \ell+2$. Thus the next order approximation to the heating rate is

\begin{equation}
\nu = 1-\frac{1}{1+\frac{1}{2C}\ln\left(\frac{1+p_2\mu^2\tau^2+p_4\mu^4\tau^4}{1+q_2\mu^2\tau^2}\right)},
\end{equation}
where the coefficients can be determined by equating the Taylor series of this function to the viscous approximation of the heating rate. Notice that, because we have three unknowns here, we must expand the viscous limit to sixth order. Similarly, the next highest order will have five unknowns, meaning that we need to expand the viscous limit to tenth order, etc.

\begin{figure}[htbp] 
   \centering
   \includegraphics[width=0.5\textwidth]{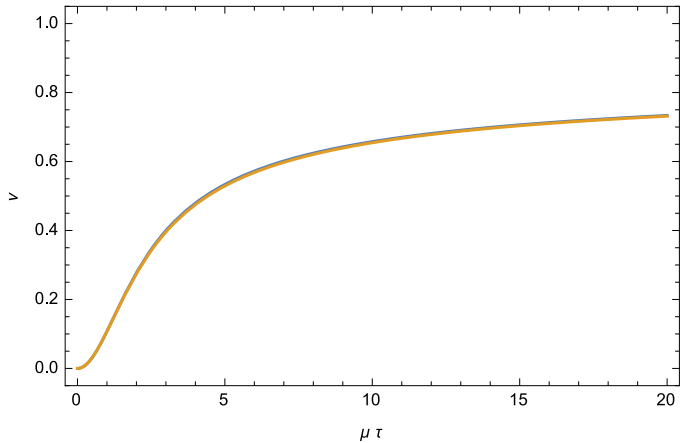} 
   \caption{The solution for the heating rate that matches the viscous limit to order $\mu^2\tau^2$ (blue curve) and to order $\mu^6\tau^6$ (yellow curve). As is apparent, the two are nearly indistinguishable, showing that the general solution for the heating rate converges rapidly to a unique solution.}
   \label{fig:nuofm}
\end{figure}

The values of the coefficients $p_n$ and $q_n$ will depend on where we truncate the viscous approximation to the heating rate, and thus our higher-order interpolated solutions for $\nu$ will differ from those at lower orders. However, we find that the overall solution appears to converge very rapidly to a unique solution, as depicted by Figure \ref{fig:nuofm}. This figure shows the lowest-order interpolated solution (i.e., equation \ref{nueq2}; blue curve) and the solution accurate to sixth order (yellow curve), both with $C = 0.812$. We see by eye that there are only very small differences between these heating rates. Therefore, we can, to a very high degree of accuracy, neglect the higher-order viscous corrections and take equation \eqref{nueq2} to be the heating rate that correctly reproduces both the viscous and streaming limits of radiation propagation and interpolates well between these extremes. Using the fact that $C \simeq 0.812$, the heating rate that preserves the self-similarity of the flow is

\begin{equation}
\nu = 1-\frac{1}{1+0.616\ln\left(1+0.217\mu^2\tau^2\right)}. \label{nuex}
\end{equation}

\section{Energy-momentum tensor}
\subsection{Comoving frame}
The energy-momentum tensor of the radiation field is \citep{mih84}

\begin{equation}
R^{\mu\nu} = \int{}f\,k^{\mu}k^{\nu}\frac{d^3k}{k^{0}}.
\end{equation}
We can evaluate this tensor at the origin by recalling that the distribution function is given by equation \eqref{fo}, using the fact that $k^{x} = k\cos\theta$ and $k^{y} = k\sin\theta\sin\phi$, and using equation \eqref{kpk} to write $k$ in terms of $k'$ and $\theta$. Doing so then gives

\begin{equation}
R^{\bar{\mu}\bar{\nu}} = \frac{e'}{4\pi}\int{}S^{\bar{\mu}\bar{\nu}}\mathscr{D}^4h^{-\frac{\nu}{m\sin\theta\sin\phi}}d\Omega, \label{Rex}
\end{equation}
where

\begin{widetext}
\begin{equation}
S^{\bar{\mu}\bar{\nu}} = \\
\left(
\begin{array}{cccc}
1 & \sin\theta\cos\phi & \sin\theta\sin\phi & \cos\theta \\
\sin\theta\cos\phi & \sin^2\theta\cos^2\phi & \sin^2\theta\sin\phi\cos\phi & \sin\theta\cos\theta\cos\phi \\
\sin\theta\sin\phi & \sin^2\theta\sin\phi\cos\phi & \sin^2\theta\sin^2\phi & \sin\theta\cos\theta\sin\phi \\
\cos\theta & \sin\theta\cos\theta\cos\phi & \sin\theta\cos\theta\sin\phi & \cos^2\theta
\end{array} \right).
\end{equation}
\end{widetext}

We have placed bars on these tensors because they are evaluated in the comoving frame of the gas parcel, and equation \eqref{Rex}, therefore, represents the \emph{comoving} radiation energy-momentum tensor. 

The heating rate $\nu$ is constructed such that $R^{\bar{0}\bar{0}} = e'$, and the symmetry of the integrand means that the energy fluxes $R^{\bar{0}\bar{z}}$, $R^{\bar{0}\bar{y}}$, and $R^{\bar{0}\bar{x}}$; and the stresses $R^{\bar{y}\bar{x}}$ and $R^{\bar{z}\bar{x}}$; are zero. It is also straightforward to show that the $x$-component of the pressure is equal to $R^{\bar{x}\bar{x}} = R^{\bar{0}\bar{0}} - R^{\bar{z}\bar{z}}-R^{\bar{y}\bar{y}}$. The only non-zero and non-trivial components of the stress tensor are therefore the pressures, $R^{\bar{z}\bar{z}}$ and $R^{\bar{y}\bar{y}}$, and the shear stress $R^{\bar{y}\bar{z}}$. 

\begin{figure}[htbp] 
   \centering
   \includegraphics[width=0.5\textwidth]{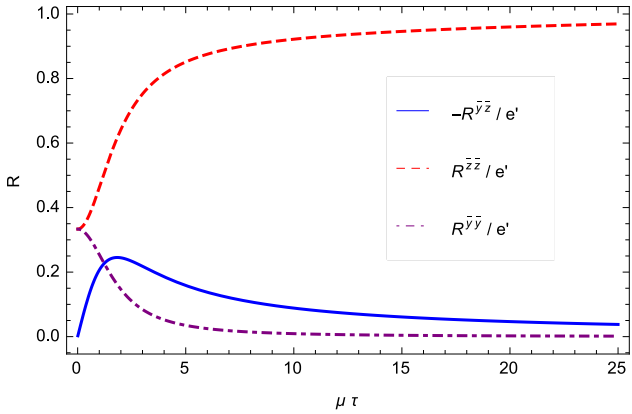} 
   \caption{The absolute value of the shear stress, $-R^{\bar{y}\bar{z}}$ (blue, solid curve), the $z$-component of the pressure, $R^{\bar{z}\bar{z}}$ (red, dashed curve), and the $y$-component of the pressure, $R^{\bar{y}\bar{y}}$ (purple, dot-dashed curve), all normalized by the energy density $e'$, as functions of the quantity $\mu\tau$. In the viscous limit, when the shear across the photon mean free path is small, so $\mu \ll 1$, we recover $R^{\bar{y}\bar{z}} = 0$ and $R^{\bar{y}\bar{y}} = R^{\bar{z}\bar{z}} = e'/3$, which is what we expect. As the shear increases, we find that the $z$-momentum flux approaches the energy density, while the shear stress and the $y$-momentum flux approach zero.}
   \label{fig:Rplots}
\end{figure}

Figure \ref{fig:Rplots} shows these three quantities -- the absolute value of the shear stress (solid, blue curve), the $z$-momentum flux (the $z$-component of the pressure; dashed, red curve) and the $y$-momentum flux (the $y$-component of the pressure; dot-dashed, purple curve) all normalized by the energy density $e'$ -- as functions of $\mu\tau$. When the shear across the photon mean free path is small, so $\mu \ll 1$, we expect the radiation field to reduce to the optically thick limit. We see from Fig.~\ref{fig:Rplots} that this consistency check is met: the shear stress approaches zero, and the momentum fluxes are $R^{\bar{z}\bar{z}} = R^{\bar{y}\bar{y}} = e'/3$ -- the value associated with an isotropic photon gas. As $\mu$ increases, we see that the shear stress initially rises in a linear fashion, reaches a peak value of $R^{\bar{y}\bar{z}} \simeq 0.25$ at $\mu\tau \simeq 1.8$, and decays asymptotically as $R^{\bar{y}\bar{z}} \propto 1/(\mu\tau)$; the $y$-component of the pressure falls off rather steeply with the shear, being well-approximated by $R^{\bar{y}\bar{y}} \simeq 1/(\mu\tau)^2$ for $\mu \gtrsim 1$; and the $z$-component of the pressure, while it rises steeply initially, eventually levels off to $R^{\bar{z}\bar{z}} = e'$ for $\mu\tau \gg 1$. 

The solutions for these quantities were obtained by numerically integrating equation \eqref{Rex} (with equation \eqref{nuex} for the heating rate), and therefore have a complicated dependence on the quantity $\mu\tau$. However, we find that each can be fit well by fairly simple functions. In particular, we find that

\begin{equation}
R^{\bar{y}\bar{z}} = -\frac{4}{15}\frac{\mu\tau\,e'}{1+\frac{3}{10}\mu^2\tau^2}, \label{Ryzapp}
\end{equation}
\begin{equation}
R^{\bar{z}\bar{z}} = \left(1-\frac{2}{3}\frac{\frac{10}{9}\tanh\left(\frac{9}{10}\mu\tau\right)}{\mu\tau}\right)e', \label{Rzzapp}
\end{equation}
\begin{equation}
R^{\bar{y}\bar{y}} = \frac{1}{3}\frac{e'}{1+\frac{1}{3}\mu^2\tau^2}. \label{Ryyapp}
\end{equation}

All of the properties of the radiation field that we have investigated thus far have been functions of the combination $\mu\tau$, thereby rendering the exact value of $\tau$ relatively unimportant. Equation \eqref{Ryzapp}, however, gives us one means to determine its value: in the viscous limit, this equation gives

\begin{equation}
R^{\bar{y}\bar{z}} = -\frac{4}{15}\mu\tau + \mathcal{O}(\mu^3\tau^3), \label{Ryzvis}
\end{equation}
which is, self-consistently, the relation we determine by Taylor expanding equation \eqref{Rex} to first order in $\mu\tau$. On the other hand, recent investigations of the equations of radiation hydrodynamics in the viscous limit have shown that (\citealt{cou14b}; see also \citealt{bla85})

\begin{equation}
R^{\bar{y}\bar{z}} = -\frac{8}{27}\mu. \label{Ryzapp1}
\end{equation}
Comparing this expression with equation \eqref{Ryzvis}, we see that the value of $\tau$ that ensures that our self-similar approach is consistent with the equations of radiation hydrodynamics in the viscous limit is

\begin{equation}
\tau_{} = \frac{10}{9}.
\end{equation}
However, it should be noted that \citet{cou14b} used a dipole scattering kernel to treat the interactions between the photons and the electrons in the gas. If one uses a spherical kernel, which is more appropriate for our analysis here (as we assumed that the distribution function was isotropic in the rest frame of the $\tau \simeq 1 $ surface; see Section 4), then one finds \citep{wei71}

\begin{equation}
R^{\bar{y}\bar{z}} = -\frac{4}{15}\mu.
\end{equation}
If we adopt this expression for the viscous limit of the radiation energy-momentum tensor, then we see that the appropriate optical depth is

\begin{equation}
\tau = 1.
\end{equation}

\subsection{Covariant formulation}
Equation \eqref{Rex} gives the radiation energy-momentum tensor in the comoving frame of the fluid. However, the equations of radiation hydrodynamics, which govern the interaction between the scatterers and the radiation field, are given by

\begin{equation}
\nabla_{\alpha}\left(T^{\alpha\beta}+R^{\alpha\beta}\right) = 0,
\end{equation}
where $\nabla_\alpha$ is the covariant derivative and $T^{\alpha\beta}$ is the energy-momentum tensor of the scatterers. It is therefore necessary to differentiate the energy-momentum tensor of the radiation field, which is tantamount to evaluating this tensor at different locations within the fluid, and hence just knowing its form at the origin is insufficient. However, from the self-similar nature of the shear flow, we know that at any comoving point the radiation field must have the form given by \eqref{Rex}. Since the comoving frame can be obtained by making a local Lorentz transformation, we therefore have

\begin{equation}
R^{\alpha\beta} = \Lambda^{\alpha}_{\,\,\bar{\alpha}}\Lambda^{\beta}_{\,\,\bar{\beta}}R^{\bar{\alpha}\bar{\beta}}.
\end{equation}
where

\begin{equation}
\Lambda^{\bar{\mu}}_{\,\,\mu} = \left(
\begin{array}{cccc}
\Gamma & 0 & 0 & \Gamma{v} \\
0 & 1 & 0 & 0 \\
0 & 0 & 1 & 0 \\
\Gamma{v} & 0 & 0 & \Gamma
\end{array} \right). 
\end{equation}
Carrying out the multiplication, this gives

\begin{equation}
R^{\alpha\beta} = \left(
\begin{array}{cccc}
\Gamma^2\left(e'+v^2R^{\bar{z}\bar{z}}\right) & 0 & \Gamma{v}R^{\bar{y}\bar{z}} & \Gamma^2v\left(e'+R^{\bar{z}\bar{z}}\right) \\
0 & R^{\bar{x}\bar{x}} & 0 & 0 \\
\Gamma{v}R^{\bar{y}\bar{z}} & 0 & R^{\bar{y}\bar{y}} & \Gamma{}R^{\bar{y}\bar{z}} \\
\Gamma^2v\left(e'+R^{\bar{z}\bar{z}}\right) & 0 & \Gamma{}R^{\bar{y}\bar{z}} & \Gamma^2\left(v^2e'+R^{\bar{z}\bar{z}}\right)
\end{array} \right).
\end{equation}
We can show that this matrix can be written

\begin{multline}
R^{\alpha\beta} = e'U^{\alpha}U^{\beta}+R^{\bar{z}\bar{z}}\Pi^{\alpha\beta} \\ +\frac{\lambda'}{\mu}R^{\bar{y}\bar{z}}\Pi^{\alpha\sigma}\Pi^{\beta\rho}\left(\nabla_{\sigma}U_{\rho}+\nabla_{\rho}U_{\sigma}\right)
+\Delta{P}^{\alpha\beta}, \label{Ralbet}
\end{multline}
where

\begin{equation}
\Pi^{\alpha\beta} = U^{\alpha}U^{\beta}+g^{\alpha\beta}
\end{equation}
is the projection tensor and we defined

\begin{equation}
\Delta{P}^{\alpha\beta} \equiv \left(
\begin{array}{cccc}
0 & 0 & 0 & 0 \\
0 & R^{\bar{x}\bar{x}}-R^{\bar{z}\bar{z}} & 0 & 0 \\
0 & 0 & R^{\bar{y}\bar{y}}-R^{\bar{z}\bar{z}} & 0 \\
0 & 0 & 0 & 0
\end{array} \right),
\end{equation}
which measures the degree of pressure anisotropy exhibited by the radiation field. 

When the amount of shear present in the flow is vanishingly small, the components of the comoving radiation energy-momentum tensor reduce to $R^{\bar{y}\bar{z}} \simeq 0$, $R^{\bar{z}\bar{z}} \simeq R^{\bar{y}\bar{y}} \simeq R^{\bar{x}\bar{x}} \simeq e'/3$. Using these expressions in equation \eqref{Ralbet}, we see that we self-consistently recover the result for the energy-momentum tensor of an isotropic, relativistic fluid:

\begin{equation}
R^{\alpha\beta} = e'U^{\alpha}U^{\beta}+\frac{1}{3}e'\Pi^{\alpha\beta}.
\end{equation}
If we keep first-order shear corrections to the comoving energy-momentum tensor, so we maintain pressure anisotropy but have $R^{\bar{y}\bar{z}} = -4\mu\tau{e'}/15$, then we find

\begin{equation}
R^{\alpha\beta} = e'U^{\alpha}U^{\beta}+\frac{1}{3}e'\Pi^{\alpha\beta}-\frac{4e'\tau}{15\rho'\kappa}\Pi^{\alpha\sigma}\Pi^{\beta\rho}\left(\nabla_{\sigma}U_{\rho}+\nabla_{\rho}U_{\sigma}\right).
\end{equation}
Comparing this expression to equation (37) of \citet{cou14b}, we see that this approach to analyzing relativistic shear is in agreement with the viscous equations of radiation hydrodynamics for divergenceless flow ($\nabla_{\mu}U^{\mu} = 0$) if we adopt $\tau = 10/9$. 

On the other hand, when the amount of shear becomes very large across the mean free path of the photon ($\mu \gg 1$), we have $R^{\bar{x}\bar{x}} \simeq R^{\bar{y}\bar{y}} \simeq R^{\bar{y}\bar{z}} \simeq 0$ and $R^{\bar{z}\bar{z}} \simeq e'$. Furthermore, since the flow is highly relativistic in this case even in the immediate vicinity of any comoving gas parcel, we have $\Gamma^2 \simeq \Gamma^2v^2$ and it therefore follows from equation \eqref{Ralbet} that

\begin{equation}
R^{\alpha\beta} = 2e'U^{\alpha}U^{\beta}.
\end{equation}

When the shear is between these two limits, relatively simple covariant expressions for the energy-momentum tensor of the radiation field are not able to be obtained. However, we note that equation \eqref{Ralbet} is valid for arbitrary $\mu$, but the full $\mu$-dependence of the comoving stress tensor must be incorporated. Furthermore, by using equation \eqref{gammacosh}, we can show that the shear parameter $\mu$ can be expressed as a covariant scalar via:

\begin{equation}
\mu^2 = \left(\lambda'\right)^2\Pi^{\mu\sigma}\left(\nabla_{\mu}U_{\nu}\right)\left(\nabla_{\sigma}U^{\nu}\right). \label{muco}
\end{equation}
The right-hand side of this equation can be interpreted as the relativistic ``square'' of the shear over the mean free path of the photon. 

\section{Summary and Discussion}
In this paper we analyzed how a radiation field responds to regions of intense, relativistic shear, which likely arise in extreme astrophysical environments such as collapsars \citep{mac99} and tidal disruption events \citep{cou14a}. We considered a two-dimensional, planar shear flow, with the motion along the $z$-direction and the variation in that motion along the $y$-direction, in which the fluid appears identical at every comoving point. We demonstrated that this self-similar assumption requires the velocity profile of the fluid to have the form $\Gamma = \cosh(\mu\tau'_y)$ with $\mu$ a constant and $\tau'_y = \rho'\kappa\,{y}$ (and the comoving density $\rho'$ is a constant to preserve the self-similarity of the flow). Using this velocity field, we determined the $\tau \simeq 1$ surface -- the location within the fluid where the integrated optical depth along the line of sight equals roughly one -- which is a complicated function of viewing angle and shear owing to relativistic Doppler beaming (see Figure \ref{fig:tausurfs} and equation \ref{rsurf}). 

Using the structure of the $\tau \simeq 1$ surface and the assumption that photons are scattered isotropically in the rest frame of the scatterer, we showed that this type of shear flow exactly conserves photon number if the distribution function of the radiation field is given by equation \eqref{dist1}. We also demonstrated that, if the energy density is to be uniform throughout the shear layer, which must be true if the scattering is elastic and the self-similarity of the fluid is upheld, then the radiation field must be heating up exponentially with a heating rate that is given by the solution to an integro-algebraic equation (see equations \ref{nueq1} and \ref{Ieq}). The solutions to this equation were determined in the limits of small and large shear, and an approximate heating rate that interpolates between these limits and almost exactly satisfies the integro-algebraic equation (see Figure \ref{fig:Iofm}) was found (equation \ref{nuex}).

Finally, we constructed the comoving energy-momentum tensor of the radiation field, the only non-zero and non-trivial components of which were the shear, $R^{\bar{y}\bar{z}}$, and the $z$- and $y$-components of the pressure, $R^{\bar{z}\bar{z}}$ and $R^{\bar{y}\bar{y}}$. These quantities were shown to smoothly transition from their viscous (small shear) to their streaming (large shear) limits, the former characterized by a shear stress that varies linearly with the shear (i.e., Newtonian in nature) and isotropic pressure, the latter portraying vanishing shear stress and highly anisotropic pressure ($R^{\bar{y}\bar{y}} \simeq R^{\bar{x}\bar{x}} \simeq 0$, $R^{\bar{z}\bar{z}} \simeq e'$; see Figure \ref{fig:Rplots}). We showed that $R^{\bar{y}\bar{z}}$, $R^{\bar{y}\bar{y}}$, and $R^{\bar{z}\bar{z}}$ were very well-fit by approximate, analytic functions (see equations \ref{Ryzapp} -- \ref{Ryyapp}), and found that the value of the optical depth must be equal to $\tau = 10/9$ if the viscous limits of our equations match those pursued by other authors (though a value of $\tau = 1$ matches the results if an isotropic scattering kernel is used in the Boltzmann equation). By using the self-similarity of the fluid, we were able to construct the form of the energy-momentum tensor at any point within the flow, showing that the result agreed with the isotropic and viscous limits.

Figure \ref{fig:Rplots}, together with equations \eqref{Ryzapp} -- \eqref{Ryyapp}, is perhaps the most important result of this investigation: this figure shows how the radiation field in a relativistically-moving plasma transitions from the viscous to the streaming limit. Interestingly, the shear stress, $R^{\bar{y}\bar{z}}$, does not grow to arbitrarily-large values as the shear increases (as one might predict from its linear growth at small $\mu$), but reaches a maximum when $\mu \simeq 1.8$ and thereafter decays approximately as $1/\mu$. The origin of this behavior can be understood as follows: the shear stress $R^{\bar{y}\bar{z}}$ gives the amount of $y$-momentum transferred in the $z$-direction. As the shear starts to increase, the structure of the $\tau = 1$ surface deviates from a sphere at $r = \lambda'$, allowing photons with negative $y$-momentum to be transferred in the positive $z$-direction (and vice versa; see Figure \ref{fig:tausurfs}) and generating the stress. When the shear becomes very large, however, the surface becomes increasingly aligned with the $z$-axis as a consequence of relativistic Doppler beaming, meaning that only photons possessing a small amount of transverse momentum are perceived in the comoving frame. The radiation field thus becomes highly beamed in the direction of motion of the fluid, thus inhibiting the transfer of transverse momentum to the scatterers. This behavior also shows that the effects of radiation drag can actually be quenched in an optically thin, relativistic shear layer, ultimately due to the fact that the perceived radiation field adapts to the presence of the shear itself.

Figure \ref{fig:Rplots} also demonstrates that the $z$-component of the pressure approaches the energy density as the shear becomes very large, as one would suspect for a highly beamed source. In fact, the red, dashed curve in that figure represents the Eddington factor $f(\mu\tau)$, which relates the $z$-component of the pressure to the energy density via $R^{\bar{z}\bar{z}} = f(\mu\tau)\,e'$. Investigating equation \eqref{Rzzapp}, we thus see that shear in a scattering medium generates an effective Eddington factor that can be well approximated by

\begin{equation}
f(\mu) = 1-\frac{2}{3}\frac{\tanh(\mu)}{\mu}, \label{fmu}
\end{equation}
where we have set $\tau = 10/9$ (which, as we demonstrated above, is the value we expect if this method is to reduce to the viscous limit when the scattering is modeled by a dipole kernel). 

Our approach adopted a very specific form for the velocity of the fluid within the shear layer. As we argued above, this velocity has the property that the fluid looks the same at every comoving point, meaning that such a shear layer may develop naturally in regions where the fluid has ``lost memory'' of the boundary conditions or initial conditions. In further support of this notion, we can show that this exact velocity profile develops in the treatment of radiation-viscous boundary layers when the flow becomes ultrarelativistic, and we refer the reader to the Appendix for a demonstration of this fact. 

The investigation we have undertaken here is related to the radiation drag limit of the interaction between photons and matter, where the radiation field is considered a constant background that is unaltered by the scattering processes that take place. On the other hand, here we have accounted for the evolution of the radiation field -- it is heated, exponentially so, by the shear present in the flow. However, we have ignored the back reaction that this heating (and the viscous stress) must have on the flow itself; similar to the fictitious entity that maintains the isotropy of the radiation field in the radiation drag limit, there must be some external force in our model that maintains the shear profile of the self-similar flow. It is the work done by this force that ultimately heats the radiation field.

We also reiterate the point made in the Introduction and in Section 2: owing to its self-similar nature, certain \emph{regions} of  astrophysical plasmas may naturally conform to the shear profile given by equation \eqref{gammacosh}. In such cases, the amount of shear present in the flow is maintained by the non-self-similar aspects of the problem, being, e.g., the boundary conditions or gradients along the direction of motion of the fluid. We verified this notion by showing, in the Appendix, that the velocity profile of the two-stream, radiation-viscous boundary layer manifestly yields the self-similar form given by equation \eqref{gammacosh}. In this case it is the boundary conditions -- that the fluid match the speed of the ``jet'' in one limit and approach the static envelope in another limit -- that provide and maintain the shear of the flow. Nevertheless, far away from these boundaries the specifics of those constraints are lost and the velocity transitions to the self-similar flow field of equation \eqref{gammacosh}.

In our treatment, the comoving density, $\rho'$, was considered independent of position and time, which is fundamental to the assumption of self-similarity within the boundary layer. However, it is possible to generalize this aspect of the problem by simply keeping the integral expression for the variable $\tau'_y$, which appears in equation \eqref{tauy}. Doing so, we can construct the $\tau \simeq 1$ surface in an identical manner to what was done in Section 3, integrate the expression exactly, and solve for the radial position of the surface as a function of angle, which gives

\begin{equation}
\int_0^{r\sin\theta\sin\phi}{\rho'\kappa\,d\tilde{y}} = \tau'_y = \frac{1}{\mu\sin\theta\sin\phi}\ln(h),
\end{equation}
where $h$ is still given by equation \eqref{heq}. What this finding demonstrates is that, since the velocity is only a function of $\mu\tau'_y$ (equation \ref{gammacosh}), the same vector transformations to obtain the comoving properties of the radiation on the $\tau = 1$ surface hold, and thus the particle number is still exactly conserved if we make the same ansatz for the distribution function, equation \eqref{dist1}. However, the heating rate will \emph{not}, in general, be the same, as this property of the radiation field depends on the retarded time between the comoving frame and the $\tau = 1$ surface. We therefore must assume a specific form for $\rho'$ to calculate $\nu$.

\citet{fuk08} followed a similar procedure to what we outlined here for calculating the properties of the radiation field: he constructed a surface in the comoving frame of the fluid that satisfied $\tau = 1$ (his ``one-tau photo-oval''), and he calculated the properties of the radiation field in the comoving frame based on the appearance of the field on the $\tau = 1$ surface. However, his treatment assumed that the fluid satisfied $\mathbf{v} = v(z)\hat{z}$, i.e., one-dimensional flow in which there is no transverse shear. He also let the comoving velocity field be $v(z) = v_0+(dv/dz)(z-z_0)$ and similarly for other fluid quantities, meaning that his results are only valid when the shear over the mean free path is small. The heating of the radiation field due to relativistic time delays was also ignored in his model, which was an essential aspect of our formulation.

Most radiation hydrodynamics codes employ a closure scheme -- either some variant of M1 or flux-limited diffusion -- to calculate some of the moments (namely the shear stresses and the pressures) of the radiation energy-momentum tensor. However, there are a few authors who have opted to directly solve the radiative transfer equation \citep{jia14, rya15} alongside the equations of radiation hydrodynamics (or magnetohydrodynamics), thereby directly computing the moments of the radiation field and coupling them to the equations of motion (and vice versa). While the methods we have outlined here were not directly based on the relativistic transfer equation, many of the properties of the fluid and the radiation field -- the self-similar appearance of the flow, the consistent transition between the optically thick and thin limits, the conservation of energy and particle number -- should also result from an analysis of the Boltzmann equation. We therefore feel that equations \eqref{Ryzapp} -- \eqref{Ryyapp} (or something close to them) should arise from an investigation of the relativistic transfer equation, and hence the comoving components of the radiation stress tensor obtained in this paper can be considered as a test for relativistic radiation-MHD solvers.

Our results concerning the properties of the radiation field were rigorously obtained only for flow in which the velocity varies as $\Gamma = \cosh(\mu\tau'_y)$. However, investigating equation \eqref{Ralbet}, we see that the form for the stress tensor that we derived only depends on this assumption through the inclusion of the parameter $\mu$ (and its assumed-constant nature). Furthermore, if we recall that $\mu$ can be characterized as a covariant scalar via equation \eqref{muco} that also \emph{does not depend on the explicit form for the velocity profile}, then we can plausibly interpret equation \eqref{Ralbet} as the radiation stress tensor that is valid for more general, but divergenceless, flows when the shear becomes large. If we additionally want to include flows that contain non-zero divergence, then comparison between our equation \eqref{Ralbet} and equation (37) of \citet{cou14b} suggests that the generalization of the viscous stress tensor to arbitrary shear is

\begin{equation}
R^{\alpha\beta} = -\frac{8}{27}\frac{e'}{\rho'\kappa}\frac{\Pi^{\alpha\sigma}\Pi^{\beta\rho}\left(\nabla_{\sigma}U_{\rho}+\nabla_{\rho}U_{\sigma}-\frac{2}{3}g_{\sigma\rho}\nabla_{\nu}U^{\nu}\right)}{1+\frac{10}{27}\frac{1}{(\rho'\kappa)^2}\Pi^{\mu\sigma}(\nabla_{\mu}U_{\nu})(\nabla_{\sigma}U^{\nu})}.
\end{equation}
We plan to investigate the validity of this relation and explore its uses in analyzing simple shear flows in a future paper.

\acknowledgements
This work was supported in part by NASA Astrophysics Theory Program grants NNX14AB37G, NSF grant AST-1411879, and NASA's Fermi Guest Investigator Program. We thank Charles Gammie for useful comments, particularly for suggesting the use of these solutions as a test problem for current radiation-MHD codes.

\appendix

\section{Viscous shear layers and self-similar flow}
\citet{cou15a} used the viscous equations of radiation hydrodynamics to analyze the two-stream boundary layer, which treats the transition between a relativistic jet and a surrounding envelope as confined to a thin (on the order of the square root of the mean free path of the photon) layer and considers the jet and the ambient medium as two distinct fluids. In this problem, the velocity of the fluid is primarily along the $z$-direction and the variation in the properties of the fluid occurs predominantly along the $y$-direction, meaning that the geometry of the problem is identical to that established in the preceding sections. Following the procedure outlined in \citet{cou15a}, we define the four-velocity in the $z$-direction as 

\begin{equation}
\Gamma{v}_z = \Gamma_jv_jf_{\xi}(\xi) \label{vss}
\end{equation}
and the comoving density as

\begin{equation}
\rho' = \rho'_0g(\xi). \label{rhoss}
\end{equation}
In these definitions, $v_j$ is the asymptotic velocity of the jet (and $\Gamma_j = (1-v_j^2)^{-1/2}$ is its Lorentz factor), $\rho'_0$ is the density of the ambient medium, $f$ and $g$ are functions of the self-similar variable 

\begin{equation}
\xi = \int_0^{y}\rho'\kappa\,dy,
\end{equation}
which we note is identical to our definition of $\tau'_y$ (see equation \ref{tauy}), and subscript $\xi$'s denote differentiation with respect to $\xi$ (i.e., $f_{\xi} = df/d\xi$, $f_{\xi\xi} = d^2f/d\xi^2$, etc.). Inserting equations \eqref{vss} and \eqref{rhoss} into the $z$-component of the momentum equation and the gas energy equation (and keeping only lowest-order terms in the boundary layer thickness; see \citet{cou15a} for details of this procedure) then yields the following two self-similar equations:

\begin{equation}
-\frac{1}{2}\left(g+\frac{4}{3}\chi\right)f\,f_{\xi\xi}+\chi\,\Gamma_j^2v_j^2\frac{gf_{\xi}\left(f_{\xi\xi}\right)^2}{1+\Gamma_j^2v_j^2\left(f_\xi\right)^2} = \chi{}\,g\,f_{\xi\xi\xi}, \label{ss1}
\end{equation}

\begin{equation}
g_{\xi}f = \frac{3}{2}\Gamma_j^2v_j^2\frac{g^2(f_{\xi\xi})^2}{1+\Gamma_j^2v_j^2(f_\xi)^2}. \label{ss2}
\end{equation}
In these equations $\chi = e'_0/\rho'_0$ is the ratio of the comoving radiation energy density in the jet to the comoving density of scatterers (they defined this quantity by $\mu$, which we avoided for obvious reasons). By direct substitution, we can show that the assumption

\begin{equation}
\Gamma{v_z} = \sinh(\mu\,\xi), \label{Gammass}
\end{equation}
where $\mu$ is an unspecified constant, exactly cancels the second term on the left-hand side of equation \eqref{ss1} with the right-hand side. Furthermore, the first term in this same equation is proportional to $1/(\Gamma_j^2v_j^2)$ so that, in the relativistic limit, this solution satisfies equation \eqref{ss1} to order $\mathcal{O}(1/\Gamma_j^2)$. 

Inserting this solution into equation \eqref{ss2}, the equation for the density becomes

\begin{equation}
\frac{g_{\xi}}{g^2} = \frac{3}{2}\mu^3\Gamma_jv_j\frac{1}{\cosh(\mu\,\xi)}.
\end{equation}
This equation can be integrated and, redefining the density $\rho'_0$ appearing in equation as the density at $\xi = 0$, so that $g(0) = 1$, we find

\begin{equation}
g = \frac{1}{1-3\mu^2\Gamma_jv_j\arctan(\tanh(\frac{\mu\xi}{2}))}. \label{gss2}
\end{equation}
Since $\tanh(\pm\infty) = \pm1$, this expression demonstrates that the comoving density varies between $(1\pm3\pi\mu^2\Gamma_jv_j/4)^{-1}$. Thus, as long as the inequality $\mu \lesssim 1/\sqrt{\Gamma_jv_j}$ is satisfied, the comoving density is nearly constant in regions where the velocity varies as $\Gamma{v_z} \simeq \sinh(\mu\xi)$, which is consistent with what we assumed based on the requirement of self-similarity in Section 2 of this paper.

Equations \eqref{Gammass} and \eqref{gss2} cannot provide the solution throughout the entire two-stream boundary layer because they do not satisfy the boundary conditions (i.e., the velocity must approach zero as we proceed into the ambient medium and it must equal the jet velocity as we go into the jet). However, what we have shown is that, at any comoving point within the flow that is sufficiently far from the boundaries, the velocity and density profile do approach the self-similar forms that we assumed in our treatment here. Since the viscous equations of radiation hydrodynamics in the boundary layer limit are the same regardless of the boundary conditions, we likewise expect this velocity profile to appear in other relativistic, radiation-viscous problems (e.g., the free-streaming jet boundary layer; \citealt{cou15b}). We thus expect that the self-similar velocity profile $\Gamma{v} = \sinh(\mu\tau'_y)$ is an inherent feature of most relativistic shear flows.

\bibliographystyle{aasjournal}
\bibliography{/Users/ecogs89/Library/texmf/bibtex/bib/mybib}

\begin{thebibliography}{}
\expandafter\ifx\csname natexlab\endcsname\relax\def\natexlab#1{#1}\fi

\bibitem[{{Arav} \& {Begelman}(1992)}]{ara92}
{Arav}, N., \& {Begelman}, M.~C. 1992, \apj, 401, 125

\bibitem[{{Begelman} {et~al.}(2006){Begelman}, {Volonteri}, \& {Rees}}]{beg06}
{Begelman}, M.~C., {Volonteri}, M., \& {Rees}, M.~J. 2006, \mnras, 370, 289

\bibitem[{{Blandford} {et~al.}(1985){Blandford}, {Jaroszynski}, \&
  {Kumar}}]{bla85}
{Blandford}, R.~D., {Jaroszynski}, M., \& {Kumar}, S. 1985, \mnras, 215, 667

\bibitem[{{Coughlin} \& {Begelman}(2014{\natexlab{a}})}]{cou14a}
{Coughlin}, E.~R., \& {Begelman}, M.~C. 2014{\natexlab{a}}, \apj, 781, 82

\bibitem[{{Coughlin} \& {Begelman}(2014{\natexlab{b}})}]{cou14b}
---. 2014{\natexlab{b}}, \apj, 797, 103

\bibitem[{{Coughlin} \& {Begelman}(2015{\natexlab{a}})}]{cou15a}
---. 2015{\natexlab{a}}, \apj, 809, 1

\bibitem[{{Coughlin} \& {Begelman}(2015{\natexlab{b}})}]{cou15b}
---. 2015{\natexlab{b}}, \apj, 809, 2

\bibitem[{{Debbasch} \& {van Leeuwen}(2009)}]{deb09a}
{Debbasch}, F., \& {van Leeuwen}, W.~A. 2009, Physica A Statistical Mechanics
  and its Applications, 388, 1079

\bibitem[{{Fukue}(2008)}]{fuk08}
{Fukue}, J. 2008, \pasj, 60, 377

\bibitem[{{Jiang} {et~al.}(2014){Jiang}, {Stone}, \& {Davis}}]{jia14}
{Jiang}, Y.-F., {Stone}, J.~M., \& {Davis}, S.~W. 2014, \apj, 796, 106

\bibitem[{{Levermore}(1984)}]{lev84}
{Levermore}, C.~D. 1984, \jqsrt, 31, 149

\bibitem[{{Levermore} \& {Pomraning}(1981)}]{lev81}
{Levermore}, C.~D., \& {Pomraning}, G.~C. 1981, \apj, 248, 321

\bibitem[{{MacFadyen} \& {Woosley}(1999)}]{mac99}
{MacFadyen}, A.~I., \& {Woosley}, S.~E. 1999, \apj, 524, 262

\bibitem[{{McKinney} {et~al.}(2014){McKinney}, {Tchekhovskoy}, {Sadowski}, \&
  {Narayan}}]{mck14}
{McKinney}, J.~C., {Tchekhovskoy}, A., {Sadowski}, A., \& {Narayan}, R. 2014,
  \mnras, 441, 3177

\bibitem[{{Mihalas} \& {Mihalas}(1984)}]{mih84}
{Mihalas}, D., \& {Mihalas}, B.~W. 1984, {Foundations of radiation
  hydrodynamics}

\bibitem[{{O'Dell}(1981)}]{ode81}
{O'Dell}, S.~L. 1981, \apjl, 243, L147

\bibitem[{{Ohsuga} {et~al.}(2009){Ohsuga}, {Mineshige}, {Mori}, \&
  {Kato}}]{ohs09}
{Ohsuga}, K., {Mineshige}, S., {Mori}, M., \& {Kato}, Y. 2009, \pasj, 61, L7

\bibitem[{{Phinney}(1982)}]{phi82}
{Phinney}, E.~S. 1982, \mnras, 198, 1109

\bibitem[{{Robertson}(1937)}]{rob37}
{Robertson}, H.~P. 1937, \mnras, 97, 423

\bibitem[{{Ryan} {et~al.}(2015){Ryan}, {Dolence}, \& {Gammie}}]{rya15}
{Ryan}, B.~R., {Dolence}, J.~C., \& {Gammie}, C.~F. 2015, \apj, 807, 31

\bibitem[{{S{\c a}dowski} {et~al.}(2014){S{\c a}dowski}, {Narayan}, {McKinney},
  \& {Tchekhovskoy}}]{sad14}
{S{\c a}dowski}, A., {Narayan}, R., {McKinney}, J.~C., \& {Tchekhovskoy}, A.
  2014, \mnras, 439, 503

\bibitem[{{S{\c a}dowski} {et~al.}(2013){S{\c a}dowski}, {Narayan},
  {Tchekhovskoy}, \& {Zhu}}]{sad13}
{S{\c a}dowski}, A., {Narayan}, R., {Tchekhovskoy}, A., \& {Zhu}, Y. 2013,
  \mnras, 429, 3533

\bibitem[{{Weinberg}(1971)}]{wei71}
{Weinberg}, S. 1971, \apj, 168, 175

\bibitem[{{Woosley}(1993)}]{woo93}
{Woosley}, S.~E. 1993, \apj, 405, 273

\end{thebibliography}

\end{document}